\documentclass[11pt]{article} 

\usepackage[utf8]{inputenc} 
\usepackage{comment}

\usepackage[numbers]{natbib}
\usepackage{hyperref}
\usepackage{geometry} 
\usepackage{graphicx} 
\usepackage{booktabs}
\usepackage{array} 
\usepackage{paralist} 
\usepackage{verbatim}
\usepackage{bm} 
\usepackage{subfig} 
\usepackage{amsmath}
\usepackage{amssymb}

\usepackage{fancyhdr} 
 
\usepackage{sectsty}
\allsectionsfont{\sffamily\mdseries\upshape} 
\usepackage[nottoc,notlof,notlot]{tocbibind} \usepackage[titles,subfigure]{tocloft}

\numberwithin{equation}{section}

\title{Complex Stochastic Optimal Control Foundation of Quantum Mechanics}
\author{Vasil Yordanov \\
	v.yordanov@phys.uni-sofia.bg \\
	\textit{Faculty of Physics, Sofia University,}\\
	\textit{5 James Bourchier blvd., 1164 Sofia, Bulgaria}
}

\begin{document}
\maketitle
\abstract 
Recent studies have extended the use of the stochastic Hamilton-Jacobi-Bellman (HJB) equation to include complex variables for deriving quantum mechanical equations. 
However, these studies often assume that it is valid to apply the HJB equation directly to complex numbers, an approach that overlooks the fundamental problem of comparing complex numbers when finding optimal controls. 
This paper explores the application of the HJB equation in the context of complex variables. It provides an in-depth investigation of the stochastic movement of quantum particles within the framework of stochastic optimal control theory.
We obtain the complex diffusion coefficient in the stochastic equation of motion using the Cauchy-Riemann theorem, considering that the particle's stochastic movement is described by two perfectly correlated real and imaginary stochastic processes. 
During the development of the covariant form of the HJB equation, we demonstrate that if the temporal stochastic increments of the two processes are perfectly correlated, then the spatial stochastic increments must be perfectly anti-correlated, and vice versa. 
The diffusion coefficient we derive has a form that enables the linearization of the HJB equation.
The method for linearizing the HJB equation, along with the subsequent derivation of the Dirac equation, was developed in our previous work [V. Yordanov, Scientific Reports 14, 6507 (2024)]. These insights deepen our understanding of quantum dynamics and enhance the application of stochastic optimal control theory to quantum mechanics.

\section{Introduction}
Quantum mechanics has several interpretations, such as the Copenhagen Interpretation, Bohmian Mechanics, Stochastic Mechanics, and others, all attempting to explain the behavior of particles at the quantum level.
These interpretations diverge on fundamental questions: Is quantum mechanics deterministic or probabilistic? Does it adhere to locality or allow for non-local interactions? Which aspects of quantum mechanics can be considered real? What is the true nature of measurement? What is the bridge leading from classical dynamics to quantum mechanics?

There have been numerous experimental attempts, including those reported in~\cite{Kurtsiefer1997,Das2022,Frabboni2012,Aharonov1988,Kocsis2011,Mahler2016,Murch2013,Rossi2019,Hariri2019,Wang2022}, as well as theoretical proposals~\cite{Das2019,Das2019_2,Rafsanjani2023,Beau2024,Naidon2024}, aimed at verifying which of these interpretations are correct.

Among these interpretations, Stochastic Mechanics stands out as the central focus of this paper. 
Within Stochastic Mechanics, two primary frameworks are currently in use: the Stochastic Variational Calculus and Stochastic Optimal Control theory. Researchers using these frameworks describe the stochastic motion of particles through different combinations of coordinate and velocity representations. These include real coordinates with real velocities (such as forward and backward velocities, or current and osmotic velocities), real coordinates with complex velocities, and complex coordinates with complex velocities.


In the following Section~\ref{sec:SM_history}, we provide a brief historical overview of the development of Stochastic Mechanics, focusing on the contributions of one of the field’s most significant authors. While this historical perspective is important for understanding the broader context, it is essential to note that the current paper is particularly relevant to the works that employ complex coordinates and complex velocities, especially in the context of Stochastic Optimal Control.

This paper clarifies the specific context in which complex numbers can be appropriately used in Stochastic Optimal Control theory. We demonstrate that the formal substitution of real numbers with complex ones in the Hamilton-Jacobi-Bellman (HJB) equation is indeed justified. However, this substitution requires consideration of two correlated stochastic processes: one governing the real part and the other governing the imaginary part of the particle coordinates. It is also crucial to assume that both the Lagrangian $\mathcal{L}$ and the intermediate action $J$ can be analytically continued into the complex plane. Additionally, when minimizing the expectation value of the complex action, it is necessary to minimize both its real and imaginary parts.

Furthermore, we provide mathematical proof that the resulting equations for determining the optimal control policy are consistent with those in the real-valued framework. 
In the complex-valued case, these equations involve taking the complex derivative of the intermediate action $J$. 

During the derivation of the complex HJB equation, we establish that, due to the Cauchy-Riemann theorem and the assumption of perfect correlation between the real and imaginary stochastic processes with equal diffusion coefficients, the resulting complex diffusion coefficient is purely imaginary.
We demonstrate that in order to preserve the covariant form of the HJB equation, the perfect correlation between the real and imaginary stochastic processes must be opposite for the temporal and spatial parts of the processes. 
For example, if the temporal part of the stochastic processes is perfectly correlated, then their spatial parts should be perfectly anti-correlated (correlation is -1). 
Moreover, this diffusion coefficient allows for the linearization of the HJB equations and thus the derivation of the Dirac equation~\cite{Yordanov2024} (see Appendix~\ref{sec:linearizing_HJB} for more details).

The derivation of the complex HJB equation is based on methodologies outlined in previous research~\cite{Kappen2005,Kappen2011,Fleming2006,Yordanov2023}. 
These works, which derive the HJB equation without considering complex numbers, establish a starting point for extending the equation to incorporate complex variables, as demonstrated in this paper.

We intentionally avoid using the Einstein summation convention due to the importance of carefully considering the indices related to the diffusion coefficients.
In this work, we restrict our analysis to Minkowski spacetime by employing the Minkowski metric $\eta$. 
Extending these results to more general spacetimes would require a careful discussion of covariance in stochastic theories, which is beyond the scope of this paper.

\section{Brief History of Stochastic Mechanics}
\label{sec:SM_history}
Stochastic Mechanics employs stochastic processes to provide a probabilistic explanation of quantum mechanics. 
The origins of this theory can be traced back to the works of Fürth~\cite{Furth1933, Peliti2020} and Fényes~\cite{Fenyes1952}, but it was Nelson~\cite{Nelson1966, Nelson1985} who is widely regarded as the father of Stochastic Mechanics. 

Nelson~\cite{Nelson1966} invented a Newtonian derivation of the Schrödinger equation using the theory of Brownian motion. 
In his theory, he assumed that any particle of mass $m$ undergoes constant Brownian motion with a diffusion coefficient $\sigma^2 = \frac{\hbar}{2m}$. 
More specifically, Nelson introduced the concepts of mean forward velocity $v_+$ and mean backward velocity $v_-$, defined as the mean forward and backward in time derivatives of the particle’s trajectory, respectively. 
These velocities are associated with two stochastic processes: $dx = v_+ dt + dw_+$ and $dx = v_- dt + dw_-$, where $w_+$ is a Wiener process with increments that are independent of the past values of $x$, and $w_-$ is another Wiener process whose increments are independent of the future values of $x$.
Nelson further defined the current velocity of the particle as $v = \frac{1}{2}(v_+ + v_-)$ and the osmotic velocity as $u = \frac{1}{2}(v_+ - v_-)$, terms that were widely used by subsequent researchers in their works.

Yasue~\cite{Yasue1980} formulated a Variational Principle for Stochastic Mechanics using the same stochastic processes introduced by Nelson. 
Additionally, he introduced Stochastic Optimal Control principles~\cite{Yasue1981}, where he constructed the stochastic Lagrangian $L(x, v_+, v_-)$ from a classical Lagrangians $L_c(x, v_+)$ and $L_c(x, v_-)$, defining it as: $L(x, v_+, v_-)=\frac{1}{2}[L_c(x,v_+) + L_c(x,v_-)]$.

Later, Guerra and Morato~\cite{Guerra1981, Guerra1983} use an approach similar to Yasue and further develop the Stochastic Optimal Control of quantum mechanics.
In their research they constructed the stochastic Lagrangian $L(x, v_+, v_-)$ from a classical Lagrangian $L_c(x,v_+)$ as: $L(x, v_+, v_-)=L_c(x,v_+)$.

Papiez~\cite{Papiez1981, Papiez1982} was among the first authors to apply Complex Stochastic Optimal Control theory and to introduce the concept of complex-valued stochastic processes and an imaginary diffusion coefficient $\sigma^2 = -\frac{i \hbar}{2 m}$. In his work the Lagrangian is complex valued, although his works lacked rigorous mathematical methods. 

Pavon has worked in both the Stochastic Variational approach~\cite{Pavon1995b, Pavon1996} and the Stochastic Optimal Control approach~\cite{Pavon1995}. 
He proposed the idea of combining the forward and backward velocities into a single complex velocity, which he termed ''quantum velocity``: $v_q = v - i u$, where $v$ is the current velocity and $u$ is the osmotic velocity. 
In his work, the complex-valued Lagrangian is defined on $\mathbb{R}^3 \times \mathbb{C}^3$, and he considers $\mathbb{C}^3$-valued stochastic processes with real coordinates in the stochastic equation of motion given by $d q = [v - i u] dt + d w$, where $d w=\frac{1}{2} \sqrt{\frac{\hbar}{m}} [(1-i) d w_+ + (1+i) d w_-]$.
He constructed the stochastic Lagrangian $L(x, v_+, v_-)$ from a classical Lagrangian $L_c(x,v_q)$ as: $L(x, v_+, v_-)= L_c(x,v_q)$.

The works of Kuipers~\cite{Kuipers2022,Kuipers2023} fall into the Stochastic Variational Calculus' category. 
In some works, like~\cite{Kuipers2022}, he uses complex position and velocity, while in others~\cite{Kuipers2023a}, he employs real position and complex velocity.
Kuipers is the first author to provide a physical interpretation of the imaginary position of the particle as the position of an associated particle in the background field.
The diffusion coefficient used in his works is $\sigma^2=\frac{i \hbar}{m}$.

Lindgren and Liukkonen~\cite{Lindgren2019} have used the Stochastic Optimal Control approach.
They introduce the idea that the Wick rotation or the analytic continuation to the imaginary axis in quantum mechanics and in the Schrödinger equation in particular comes from the invariant volume form on Lorentzian manifolds. The Lagrangian in their work is also complex valued.
The authors use the diffusion coefficient $\sigma^2=\frac{i \hbar}{m}$. They introduce imaginary time, but due to the feedback control policy $m u_\mu=i \nabla_\mu J$ it turns out that the coordinates and the velocities in their research are also complex numbers.
In their works they use the following stochastic equation of motion for temporal coordinates: $d(c X_0)=u_0 ds + \sqrt{R+\frac{1}{i m}} dW_0$, where $\sigma_1^2 + \sigma_2^2 + \sigma_3^2=R \in \mathbb{R}$ and for spatial coordinates $d X_i=u_i ds + \sigma_i d W_i$.

Yang and Han~\cite{Yang2013,Yang2021_Cian_Dong}, in their work, introduced the concept of Complex Quantum Mechanics, which also relies on the HJB equation. Their study explicitly defines complex stochastic equations of motion and incorporates complex functions and variables within the stochastic HJB framework. They are using as a diffusion coefficient: $\sigma^2=-\frac{i \hbar}{m}$.

In our previous research~\cite{Yordanov2024}, we derive the Dirac equation, assuming a complex equation of motion and applying the HJB equation to complex functions and variables. 
In this work the Lagrangian is a complex valued function of complex arguments, and the derived diffusion coefficient is: $\eta_{\mu \mu} \sigma^\mu \sigma^\mu=\frac{2 i \epsilon_r \hbar}{m}, \quad \mu=0,1,2,3$, where $\epsilon_r=+1$ for particles, and $\epsilon_r=-1$ for anti-particles. 
The diffusion coefficient is derived from the requirement to linearize the HJB equation. 
As a consequence of this linearization, the diffusion coefficient is twice as large as the complex diffusion coefficients used by other authors.

One can see that different authors in their research use various Lagrangians, different diffusion coefficients, and, respectively, different stochastic equations of motion. 
In the present paper, we directly complexify the stochastic equation of motion and analytically continue the Lagrangian of the system, which provides the most natural complex extension of the theory. 
However, we do not start by defining the complex diffusion coefficient as the mechanical sum: $\sigma^\mu=\sigma^\mu_x + i \sigma^\mu_y$. 
Instead, this equation arises as a consequence of the relationship between $\sigma^\mu$, $\sigma^\mu_x$ and $\sigma^\mu_y$ through the definition of the scaled complex stochastic increments.
In the end, when we postulate that the real and imaginary diffusion coefficients satisfy $\sigma^\mu_x \sigma^\mu_x = \sigma^\mu_y \sigma^\mu_y = \frac{\hbar}{m}, \quad \mu=0,1,2,3$, then the complex diffusion coefficient is derived to be: $\sigma^\mu \sigma^\mu=\frac{2 i \eta^{\mu \mu} \epsilon_r \hbar}{m}, \quad \mu=0,1,2,3$.

Current paper can be categorized to be related to the works that utilize the Stochastic Optimal Control with complex stochastic equation of motion and complex Lagrangian that is function of complex coordinates and velocities. 
As seen from the brief historical overview above, the authors that use the same approach are Papiez, Lindgren and Yang.
In their works, they have formally applied the result of real-valued Stochastic Optimal Control to that of complex numbers.
In current paper we clarify the use of Stochastic Optimal Control with complex numbers and show that it is mathematically valid to use such approach, but we will also consider in details what are the constraints of such application.

\section{Derivation of the complex stochastic HJB equation}
\label{sec:complex_HJB}
According to the first postulate of the Stochastic Optimal Control Theory of Quantum Mechanics~\cite{Nelson1985,Yordanov2024}, a particle moves like a Brownian particle in four-dimensional spacetime, influenced by an external random spacetime force. 
However, in the case of Complex Stochastic Optimal Control theory, this motion occurs within the complex plane for each of the four-coordinate components. 
Later in Section~\ref{sec:complex_stochastic_motion}, we will discuss in more detail the question of the physical interpretation of the imaginary component of the particle’s coordinate.
 
The four-coordinates of the particle are defined as complex numbers, enabling the inclusion of both real and imaginary components to fully describe its position in spacetime. We express these coordinates as follows:
\begin{equation}
\label{eq:Complex_Coordinates}
z^\mu = x^\mu + i y^\mu, \quad \mu=0,1,2,3,
\end{equation}
where $x^\mu$ and $y^\mu$ represent the real and imaginary parts, respectively, and the index $\mu$ spans the four dimensions of spacetime, aligning with the notation used in relativistic mechanics.

The velocity of the particle is similarly complexified to account for motions in both the real and imaginary dimensions of the spacetime:
\begin{equation}
\label{eq:Complex_Velocity}
w^\mu = v^\mu + i u^\mu, \quad \mu=0,1,2,3,
\end{equation}
where $v^\mu$ and $u^\mu$ denote the real and imaginary components of the four-velocity, respectively.

The complex stochastic equation of motion governing the stochastic movement of the particle is given by:
\begin{equation}
\label{eq:Complex_Stochastic_Process}
dz^\mu = w^\mu ds + \sigma^\mu dW^\mu, \quad \mu=0,1,2,3,
\end{equation}
where $w^\mu$ represents the complex velocity and $\sigma^\mu$ denotes the diffusion coefficients, with $dW^\mu$ being the increments of a Wiener process that encapsulates the stochastic nature of the particle’s trajectory. 
In the current paper, the complex Wiener process is defined as a union of two separate real Wiener processes -- one for the real and one for the imaginary coordinates of the particle. In this sense, writing the complex stochastic equation of motion~\eqref{eq:Complex_Stochastic_Process} is formal; one should always consider the following two separate stochastic equations of motion for the real and imaginary parts of the particle, represented as follows:

\begin{equation}
\label{eq:Real_Stochastic_Process}
dx^\mu=v^\mu ds+ \sigma_x^\mu dW_x^\mu, \quad \mu=0,1,2,3,
\end{equation}

\begin{equation}
\label{eq:Imaginary_Stochastic_Process}
dy^\mu=u^\mu ds+ \sigma_y^\mu dW_y^\mu, \quad \mu=0,1,2,3,
\end{equation}
where $\sigma_x^\mu$ and $\sigma_y^\mu$ denote the diffusion coefficients, with $dW_x^\mu$ and $dW_y^\mu$ representing the increments of Wiener processes. 
These stochastic terms encapsulate the stochastic nature of the particle’s trajectory, associated respectively with the stochastic dynamics of the real and imaginary components of the coordinates. 

Later, we will require the two stochastic processes to be not independent, but instead perfectly correlated or anti-correlated, either temporally or spatially. This condition is necessary to obtain the complex HJB equation. Additionally, we will demonstrate that if the real and imaginary diffusion coefficients are equal, then the complex diffusion coefficient becomes purely imaginary, allowing the linearization of the HJB equation. It is important to note that, in general, this correlation is not a requirement of the complex Wiener process.

The complex scaled stochastic increment $\sigma^\mu dW^\mu$ can be obtained by multiplying Equation \eqref{eq:Imaginary_Stochastic_Process} by the imaginary unit $i$ and adding the result to Equation \eqref{eq:Real_Stochastic_Process}: $dz^\mu = dx^\mu + i dy^\mu =(v^\mu + i u^\mu) + \sigma_x^\mu dW_x^\mu + i \sigma_y^\mu dW_y^\mu =w^\mu ds + \sigma^\mu dW^\mu$.
\begin{equation}
\label{eq:Scaled_Complex_Stochastic_Increments}
\sigma^\mu dW^\mu = \sigma^\mu_x dW^\mu_x+ i \sigma^\mu_y dW^\mu_y \quad \mu=0,1,2,3.
\end{equation}
 
This equation expresses the complex scaled stochastic increment $\sigma^\mu dW^\mu$ in terms of the real-valued diffusion coefficients $\sigma^\mu_x$ and $\sigma^\mu_y$, and their respective stochastic increments $dW^\mu_x$ and $dW^\mu_y$. 

It is important to note that this expression defines the product $\sigma^\mu dW^\mu$. In the general case, without specific information about the correlations between the stochastic processes $W^\mu_x$ and $W^\mu_y$, the individual terms $\sigma^\mu$ and $dW^\mu$ cannot be expressed solely in terms of their real counterparts $\sigma^\mu_x$, $\sigma^\mu_y$, and $dW^\mu_x$, $dW^\mu_y$, respectively.
If the stochastic processes $W^\mu_x$ and $W^\mu_y$ are perfectly correlated, it will be shown later that $\sigma^\mu = \sigma^\mu_x + i \sigma^\mu_y$, and $\sigma^\mu$  can be considered as complex diffusion coefficient, and $dW^\mu$ is a stochastic increment. In the case of perfect anti-correlation, the diffusion coefficient takes the form $\sigma^\mu = \sigma^\mu_x - i \sigma^\mu_y$ or alternatively $\sigma^\mu = - \sigma^\mu_x + i \sigma^\mu_y$.
  
According to the third postulate of Stochastic Optimal Control Theory in Quantum Mechanics, as outlined by~\cite{Lindgren2019,Yordanov2024}, Nature tries to minimize the expected value for the action, in which the particle’s velocity is consider to be a control parameter of the optimization.
Formally the complex action is defined as the minimum of the expected value of stochastic action:
\begin{equation}
\label{eq:eq_for_cost}
\begin{aligned}
S(\mathbf{z}_\mathrm{i}, \mathbf{w}(\tau_\mathrm{i} \rightarrow \tau_\mathrm{f}))=\min_{\mathbf{w}(\tau_\mathrm{i} \rightarrow \tau_\mathrm{f}) } \left< \int_{\tau_\mathrm{i}}^{\tau_\mathrm{f}} ds\, \mathcal{L}(\mathbf{z}(s), \mathbf{w}(s), s) \right>_{\mathbf{z}_\mathrm{i}}.
\end{aligned}
\end{equation}

However, this formal definition encounters difficulties because it is not possible to directly find a minimum of a complex-valued function. To address this issue, we redefine the complex action by separating its real and imaginary components, which are then independently optimized:

\begin{equation}
\label{eq:eq_for_cost}
\begin{aligned}
S(\mathbf{z}_\mathrm{i}, \mathbf{w})=&S_R(\mathbf{x}_\mathrm{i}, \mathbf{y}_\mathrm{i}, \mathbf{v}, \mathbf{u}) + i S_I(\mathbf{x}_\mathrm{i}, \mathbf{y}_\mathrm{i}, \mathbf{v}, \mathbf{u}),
\end{aligned}
\end{equation}
where the real and imaginary parts of the action are defined respectively as follows:

\begin{equation}
\label{eq:eq_for_cost}
\begin{aligned}
S_{R,I}(\mathbf{x}_\mathrm{i}, \mathbf{y}_\mathrm{i}, \mathbf{v}(\tau_\mathrm{i} \rightarrow \tau_\mathrm{f}), \mathbf{u}(\tau_\mathrm{i} \rightarrow \tau_\mathrm{f})) 
=\min_{\substack{\mathbf{v}(\tau_\mathrm{i} \rightarrow \tau_\mathrm{f}) \\ \mathbf{u}(\tau_\mathrm{i} \rightarrow \tau_\mathrm{f})}} \left< \int_{\tau_\mathrm{i}}^{\tau_\mathrm{f}} ds\, \mathcal{L}_{R,I}(\mathbf{x}(s), \mathbf{y}(s), \mathbf{v}(s),\mathbf{u}(s), s) \right>_{\mathbf{x}_\mathrm{i}, \mathbf{y}_\mathrm{i}},
\end{aligned}
\end{equation}

where $\mathcal{L}_R(\mathbf x, \mathbf y, \mathbf v, \mathbf u, s)$ and $\mathcal{L}_I(\mathbf x, \mathbf y, \mathbf v, \mathbf u, s)$ are the real and imaginary parts of the Lagrangian of the test particle. These functions depend on the real and imaginary components of the control policies $\mathbf v$ and $\mathbf u$, the and four-coordinates $\mathbf x$ and $\mathbf y$ at proper time $s$. 
The subscripts $\mathbf x_\mathrm{i}$ and $\mathbf y_\mathrm{i}$ on the expectation value indicate that the expectation is calculated over all stochastic trajectories that originate at the complex coordinate $\mathbf z_\mathrm{i}=\mathbf x_\mathrm{i} + i \mathbf y_\mathrm{i}$.

We define the complex Lagrangian of the particle as:
\begin{equation}
\label{eq:complex_lagrangian}
\begin{aligned}
\mathcal{L}(\mathbf{z}, \mathbf{w}, s)=\mathcal{L}_{R}(\mathbf{x}, \mathbf{y}, \mathbf{v},\mathbf{u}, s) + i \mathcal{L}_{I}(\mathbf{x}, \mathbf{y}, \mathbf{v}, \mathbf{u}, s),
\end{aligned}
\end{equation}
such that $\mathcal{L}(\mathbf{z}, \mathbf{w}, s)$ is the analytic continuation of the Lagrangian of the particle.

The task of optimal control theory~\cite{bellman1954} is to find the controls $\mathbf v(s)$ and $\mathbf u(s)$, $\tau_\mathrm{i} < s < \tau_\mathrm{f}$, denoted as $\mathbf v(\tau_\mathrm{i} \rightarrow \tau_\mathrm{f})$ and $\mathbf u(\tau_\mathrm{i} \rightarrow \tau_\mathrm{f})$, that minimizes the expected value of the action $S_{R,I}(\mathbf x_\mathrm{i}, \mathbf y_\mathrm{i}, \mathbf v(\tau_\mathrm{i} \rightarrow \tau_\mathrm{f}), \mathbf u(\tau_\mathrm{i} \rightarrow \tau_\mathrm{f}))$.

We introduce the optimal cost-to-go function for any intermediate proper time $\tau$, where $\tau_\mathrm{i} < \tau < \tau_\mathrm{f}$:
\begin{equation}
\begin{aligned}
\label{eq:cost_to_go_function}
& J_{R,I}(\tau, \mathbf x_\tau, \mathbf y_\tau)=\min_{\substack{\mathbf{v}(\tau \rightarrow \tau_\mathrm{f}) \\ \mathbf{u}(\tau \rightarrow \tau_\mathrm{f})}} \left< \int_{\tau}^{\tau_\mathrm{f}} ds\, \mathcal{L}_{R,I}(\mathbf{x}(s), \mathbf{y}(s), \mathbf{v}(s), \mathbf{u}(s), s) \right>_{\mathbf{x}_\tau, \mathbf{y}_\tau}.
\end{aligned}
\end{equation}

We define the complex cost-to-go function as:
\begin{equation}
J(\tau, \mathbf z) = J_R(\tau, \mathbf x, \mathbf y) + i J_I(\tau, \mathbf x, \mathbf y),
\end{equation}
assuming that it is also an analytic function in the complex plane.

By definition, the action $S_{R,I}(\mathbf x_\mathrm{i}, \mathbf v(\tau_\mathrm{i} \rightarrow \tau_\mathrm{f}), \mathbf u(\tau_\mathrm{i} \rightarrow \tau_\mathrm{f}))$ is equal to the cost-to-go function $J_{R,I}(\tau_\mathrm{i}, \mathbf x_{\tau_\mathrm{i}}, \mathbf y_{\tau_\mathrm{i}})$ at the initial proper time and spacetime coordinate:
\begin{equation}
S_{R,I}(\mathbf x_\mathrm{i}, \mathbf y_\mathrm{i}, \mathbf v(\tau_\mathrm{i} \rightarrow \tau_\mathrm{f}), \mathbf u(\tau_\mathrm{i} \rightarrow \tau_\mathrm{f}))=J_{R,I}(\tau_\mathrm{i}, \mathbf x_{\tau_\mathrm{i}}, \mathbf y_{\tau_\mathrm{i}}).
\end{equation}

We can rewrite recursive formula for $J_{R,I}(\tau, \mathbf x_\tau, \mathbf y_\tau)$ for any intermediate time $\tau'$, where $\tau < \tau' < \tau_\mathrm{f}$:
\begin{equation}
\label{eq:recursive_J}
\begin{aligned}
&J_{R,I}(\tau, \mathbf x_\tau, \mathbf y_\tau) = \\
&=\min_{\substack{\mathbf{v}(\tau \rightarrow \tau_\mathrm{f}) \\ \mathbf{u}(\tau \rightarrow \tau_\mathrm{f})}} \left< \int_\tau^{\tau'} ds\,\mathcal{L}_{R,I}(s, \mathbf x_s, \mathbf y_s, \mathbf v_s, \mathbf u_s) + \int_{\tau'}^{\tau_\mathrm{f}} ds\,\mathcal{L}_{R,I}(s, \mathbf x_s, \mathbf y_s, \mathbf v_s, \mathbf u_s) \right>_{\mathbf x_\tau, \mathbf y_\tau} \\
&=\min_{\substack{\mathbf{v}(\tau \rightarrow \tau') \\ \mathbf{u}(\tau \rightarrow \tau')}} \left< \int_\tau^{\tau'} ds\,\mathcal{L}_{R,I}(s, \mathbf x_s, \mathbf y_s, \mathbf v_s, \mathbf u_s) + \min_{\mathbf v(\tau' \rightarrow \tau_\mathrm{f} )} \left< \int_{\tau'}^{\tau_\mathrm{f}} ds\,\mathcal{L}_{R,I}(s, \mathbf x_s, \mathbf y_s, \mathbf v_s, \mathbf u_s) \right>_{\mathbf x_{\tau'},\mathbf y_{\tau'}} \right>_{\mathbf x_\tau, \mathbf y_\tau} \\
&=\min_{\substack{\mathbf{v}(\tau \rightarrow \tau') \\ \mathbf{u}(\tau \rightarrow \tau')}} \left< \int_\tau^{\tau'} ds\,\mathcal{L}_{R,I}(s, \mathbf x_s, \mathbf y_s, \mathbf v_s, \mathbf u_s) + J_{R,I}(\tau', \mathbf x_{\tau'}, \mathbf y_{\tau'}) \right>_{\mathbf x_\tau, \mathbf y_\tau}.
\end{aligned}
\end{equation}

In above equation we split the minimization over two intervals. These are not independent, because the second minimization is conditioned on the starting value $x_{\tau'}$, $y_{\tau'}$, which depends on the outcome of the first minimization~\cite{Kappen2005a,Kappen2007}. 

If $\tau'$ is a small increment of $\tau$, $\tau'=\tau + d\tau$ then:

\begin{equation}
\label{eq:J_recursivly_expressed_by_L}
J_{R,I}(\tau, \mathbf x_\tau, \mathbf y_\tau)=\min_{\substack{\mathbf{v}(\tau \rightarrow \tau + d \tau) \\ \mathbf{u}(\tau \rightarrow \tau + d \tau)}} \left<\mathcal{L}_{R,I}(\tau, \mathbf x_\tau, \mathbf y_\tau, \mathbf v_\tau, \mathbf u_\tau) d \tau + J_{R,I}(\tau+d\tau, \mathbf x_{\tau + d\tau}, \mathbf y_{\tau + d\tau}) \right>_{\mathbf x_\tau, \mathbf y_\tau}.
\end{equation}

We must perform a Taylor expansion of $J_R$ and $J_I$ in $d \mathbf x$, $d \mathbf y$ and $d\tau$. 
By applying It{\^o}'s lemma~\cite{Ito1951}, since $\left<(dx^\mu)^2 \right>=(\sigma_x^\mu)^2 d\tau$ and $\left<(dy^\mu)^2 \right>=(\sigma_y^\mu)^2 d\tau$, both of which are of order $d\tau$, we must expand up to the second order in $d \mathbf x^2$ and $d \mathbf y^2$:

\begin{equation}
\label{eq:avg_cost-to-go}
\begin{aligned}
& \left< J_{R,I}(\tau+d\tau,  \mathbf x_{\tau + d\tau}, \mathbf y_{\tau + d\tau}) \right>_{\mathbf x_\tau, \mathbf y_\tau} = \\
& = \int d \mathbf x_{\tau+d\tau} d \mathbf y_{\tau+d\tau} \mathcal N ((\mathbf x_{\tau + d\tau}, \mathbf y_{\tau + d\tau}) | (\mathbf x_\tau, \mathbf x_\tau), \sigma_x^\mu d\tau, \sigma_y^\mu d\tau) J_{R,I}(\tau + d\tau, \mathbf x_{\tau+d\tau}, \mathbf y_{\tau+d\tau}) \\
& = \int d \mathbf x_{\tau+d\tau} d \mathbf y_{\tau+d\tau} \mathcal N ((\mathbf x_{\tau+d\tau}, \mathbf y_{\tau+d\tau}) | (\mathbf x_\tau, \mathbf y_\tau), \sigma_x^\mu d\tau, \sigma_y^\mu d\tau) \times \\
& \quad \times \Big(J_{R,I}(\tau, \mathbf x_\tau, \mathbf y_\tau)+d\tau \partial_\tau J_{R,I}(\tau, \mathbf x_\tau, \mathbf y_\tau) + \sum_{\mu=0}^{3} dx^\mu \partial_{x^\mu} J_{R,I}(\tau, \mathbf x_\tau, \mathbf y_\tau) + \\
& \quad + \sum_{\mu=0}^{3} dy^\mu \partial_{y^\mu} J_{R,I}(\tau, \mathbf x_\tau, \mathbf y_\tau) +\sum_{\mu,\nu=0}^{3} dx^\mu dx^\nu \frac{1}{2} \partial_{x^\mu x^\nu} J_{R,I}(\tau, \mathbf x_\tau, \mathbf y_\tau) + \\
& \quad + \sum_{\mu,\nu=0}^{3} dy^\mu dy^\nu \frac{1}{2} \partial_{y^\mu y^\nu} J_{R,I}(\tau, \mathbf x_\tau, \mathbf y_\tau) + \sum_{\mu, \nu=0}^{3} dx^\mu dy^\nu \partial_{x^\mu y^\nu} J(\tau, \mathbf x_\tau, \mathbf y_\tau) \Big) \\
\end{aligned}
\end{equation}

From this, we obtain:
\begin{equation}
\label{eq:avg_cost-to-go}
\begin{aligned}
& \left< J_{R,I}(\tau+d\tau,  \mathbf x_{\tau + d\tau}, \mathbf y_{\tau + d\tau}) \right>_{\mathbf x_\tau, \mathbf y_\tau} = \\
& = J_{R,I}(\tau, \mathbf x_\tau, \mathbf y_\tau)+d\tau \partial_\tau J_{R,I}(\tau, \mathbf x_\tau, \mathbf y_\tau) + \sum_{\mu=0}^{3} \left<dx^\mu \right>_{\mathbf x_\tau, \mathbf y_\tau} \partial_{x^\mu} J_{R,I}(\tau, \mathbf x_\tau, \mathbf y_\tau) + \\
& \quad + \sum_{\mu=0}^{3} \left<dy^\mu \right>_{\mathbf x_\tau, \mathbf y_\tau} \partial_{y^\mu} J_{R,I}(\tau, \mathbf x_\tau, \mathbf y_\tau) + \frac{1}{2} \sum_{\mu, \nu =0}^{3} \left<dx^\nu dx^\mu \right>_{\mathbf x_\tau, \mathbf y_\tau} \partial_{x^\nu x^\mu} J_{R,I}(\tau, \mathbf x_\tau, \mathbf y_\tau) + \\
& \quad + \frac{1}{2} \sum_{\mu,\nu=0}^{3} \left<dy^\nu dy^\mu \right>_{\mathbf x_\tau, \mathbf y_\tau} \partial_{y^\nu y^\mu} J_{R,I}(\tau, \mathbf x_\tau, \mathbf y_\tau) + \sum_{\mu, \nu=0}^{3} \left<dx^\mu dy^\nu \right>_{\mathbf x_\tau, \mathbf y_\tau} \partial_{x^\mu y^\nu} J_{R,I}(\tau, \mathbf x_\tau, \mathbf y_\tau).
\end{aligned}
\end{equation}

Here $\mathcal N ((\mathbf x_{\tau + d\tau}, \mathbf y_{\tau + d\tau} )| (\mathbf x_\tau, \mathbf y_\tau), \sigma_x^\mu d\tau, \sigma_y^\mu d\tau)$ is the conditional probability starting from state $(\mathbf x_\tau, \mathbf y_\tau)$ to end up in state $(\mathbf x_{\tau+d\tau}, \mathbf y_{\tau+d\tau})$. 
The integration is over the entire spacetime for $\mathbf x$ and $\mathbf y$.

From equations~\eqref{eq:Real_Stochastic_Process} and~\eqref{eq:Imaginary_Stochastic_Process} the expected value of $dx^\mu$ and $dy^\mu$ respectively are:
\begin{equation}
\left<dx^\mu \right>_{\mathbf x_\tau, \mathbf y_\tau} = v^\mu d\tau, \quad \mu=0,1,2,3.
\end{equation}
\begin{equation}
\left<dy^\mu \right>_{\mathbf x_\tau, \mathbf y_\tau} = u^\mu d\tau, \quad \mu=0,1,2,3.
\end{equation}

We can calculate the expected values of the $\left<dx^\nu dx^\mu \right>_{\mathbf x_\tau, \mathbf y_\tau}$ using Equations~\eqref{eq:Real_Stochastic_Process} and~\eqref{eq:Imaginary_Stochastic_Process}:
\begin{equation}
\begin{aligned}
& \left< dx^\nu dx^\mu \right>_{\mathbf x_\tau, \mathbf y_\tau} = \\
& = \int d \mathbf x_{\tau + d\tau} d \mathbf y_{\tau + d\tau} \mathcal N ((\mathbf x_{\tau + d\tau}, \mathbf y_{\tau + d\tau}) | (\mathbf x_\tau, \mathbf y_\tau), \sigma_x^\mu d\tau, \sigma_y^\mu d\tau) dx^\mu dx^\nu \\
& = \int d \mathbf x_{\tau + d\tau} d \mathbf y_{\tau + d\tau} \mathcal N ((\mathbf x_{\tau + d\tau}, \mathbf y_{\tau + d\tau}) | (\mathbf x_\tau, \mathbf y_\tau), \sigma_x^\mu d\tau, \sigma_y^\mu d\tau) (v^\mu d\tau + \sigma_x^\mu d W^\mu_x) (v^\nu d\tau + \sigma_x^\mu d W^\nu_x) \\ 
& = \int d \mathbf x_{\tau + d\tau} d \mathbf y_{\tau + d\tau} \mathcal N ((\mathbf x_{\tau + d\tau}, \mathbf y_{\tau + d\tau}) | (\mathbf x_\tau, \mathbf y_\tau), \sigma_x^\mu d\tau, \sigma_y^\mu d\tau) \times\\
& \quad \times (v^\mu v^\nu d^2 \tau + v^\mu \sigma_x^\mu d\tau d W^\nu_x + \sigma_x^\mu d W^\mu_x v^\nu d\tau+ \sigma_x^\mu d W^\mu_x \sigma_x^\mu d W^\nu_x).
\end{aligned}
\end{equation}

From where we derive:
\begin{equation}
\label{eq:dx_dx_avg}
\left< dx^\nu dx^\mu \right>_{\mathbf x_\tau, \mathbf y_\tau} = 0, \quad \mu \neq \nu; \quad \left< dx^\mu dx^\mu \right>_{\mathbf x_\tau, \mathbf y_\tau} = \sigma_x^\mu \sigma_x^\mu d \tau, \quad \mu=0,1,2,3.
\end{equation}

Respectively, for the imaginary components:
\begin{equation}
\label{eq:dy_dy_avg}
\left< dy^\nu dy^\mu \right>_{\mathbf x_\tau, \mathbf y_\tau} = 0, \quad \mu \neq \nu;\quad \left< dy^\mu dy^\mu \right>_{\mathbf x_\tau, \mathbf y_\tau} = \sigma_y^\mu \sigma_y^\mu d \tau, \quad \mu=0,1,2,3.
\end{equation}

For simplicity in the following considerations, we choose the mostly negative signature convention for the metric tensor $\eta_{\mu\nu} = \text{diag}(-1, +1, +1, +1)$. 
Similar results can be obtained for the mostly positive signature convention.

We assume that the correlations between the stochastic terms in the real~\eqref{eq:Real_Stochastic_Process} and imaginary~\eqref{eq:Imaginary_Stochastic_Process} stochastic equations of motion are determined by the spacetime metric. 
Specifically, we assume perfect anti-correlation between the temporal stochastic terms $dW_x^0$ and $dW_y^0$, and perfect correlation between the spatial stochastic terms $dW_x^i$ and $dW_y^i$ (for $i=1,2,3$). 
This choice ensures that the stochastic processes are consistent with the underlying relativistic metric.

Under this assumption, we obtain the following expression for the expected value of the mixed term $dx^\mu dy^\nu$:

\begin{equation}
\label{eq:dx_dy_avg}
\left< dx^\mu dy^\nu \right>_{\mathbf x_\tau, \mathbf y_\tau} = 0, \quad \mu \neq \nu;\quad \left< dx^\mu dy^\mu \right>_{\mathbf x_\tau, \mathbf y_\tau} = \epsilon \eta^{\mu \mu} \sigma_x^\mu \sigma_y^\mu d \tau, \quad \mu=0,1,2,3.
\end{equation}

By utilizing the properties of the metric tensor, these equations can be combined into a single expression: $\left< dx^\mu dy^\nu \right>_{\mathbf x_\tau, \mathbf y_\tau} = \epsilon \eta^{\mu \nu} \sigma_x^\mu \sigma_y^\nu d \tau$.
One can easily check that $\left< dx^\mu dy^\nu \right>_{\mathbf x\tau, \mathbf y_\tau}$ is not invariant under Lorentz transformations when $\mu$ and $\nu$ are held fixed. 
However, the introduction of the metric tensor in equation~\eqref{eq:dx_dy_avg}, and hence the differing correlations between the temporal and spatial components of the stochastic processes $W_x$ and $W_y$, ensure that the mixed term in Equation~\eqref{eq:avg_cost-to-go} is Lorentz invariant. This invariance is essential for deriving the covariant form of the complex HJB equation in Section~\ref{sec:stochastic_equation_of_motion}.

Note that in Equations~\eqref{eq:dx_dx_avg} and~\eqref{eq:dy_dy_avg}, the metric tensor elements $\eta^{\mu \mu}$ do not appear. This is because these equations represent the variances of the individual stochastic increments $dx^\mu$ and $dy^\mu$, which are inherently positive quantities and do not require the metric tensor for their definition. Later, in Section~\ref{sec:stochastic_equation_of_motion}, we will see that under the hypothesis $\sigma_x^\mu \sigma_x^\mu = \sigma_y^\mu \sigma_y^\mu$, the contributions from these non-mixed terms will cancel out, ensuring that Lorentz invariance is maintained in the final HJB equation.

After substituting the above equations into equation~\eqref{eq:avg_cost-to-go}, we derive the stochastic HJB equation for the real part of our system:
\begin{equation}
\label{eq:HJB_R}
\begin{aligned}
& -\partial_\tau J_R(\tau, \mathbf x, \mathbf y)= \\
& = \min_{\mathbf v, \mathbf u} \Big( \mathcal{L}_R(\tau, \mathbf x, \mathbf y, \mathbf v, \mathbf u) + \sum_{\mu=0}^{3} v^\mu \partial_{x^\mu} J_R(\tau, \mathbf x, \mathbf y) + \sum_{\mu=0}^{3} u^\mu \partial_{y^\mu} J_R(\tau, \mathbf x, \mathbf y) \Big) + \\
& \quad + \frac{1}{2} \sum_{\mu=0}^{3} \big( \sigma_x^\mu \sigma_x^\mu \partial_{x^\mu} \partial_{x^\mu} J_R(\tau, \mathbf x, \mathbf y) + 2 \epsilon \eta^{\mu \mu} \sigma_x^\mu \sigma_y^\mu \partial_{x^\mu} \partial_{y^\mu} J_R(\tau, \mathbf x, \mathbf y) + \sigma_y^\mu \sigma_y^\mu \partial_{y^\mu} \partial_{y^\mu} J_R(\tau, \mathbf x, \mathbf y) \big),
\end{aligned}
\end{equation}

In a similar manner, the equation for the imaginary part is derived:
\begin{equation}
\label{eq:HJB_I}
\begin{aligned}
& -\partial_\tau J_I(\tau, \mathbf x, \mathbf y)= \\
& = \min_{\mathbf v, \mathbf u} \Big( \mathcal{L}_I(\tau, \mathbf x, \mathbf y, \mathbf v, \mathbf u) + \sum_{\mu=0}^{3} v^\mu \partial_{x^\mu} J_I(\tau, \mathbf x, \mathbf y) + \sum_{\mu=0}^{3} u^\mu \partial_{y^\mu} J_I(\tau, \mathbf x, \mathbf y) \Big) + \\
& \quad + \frac{1}{2} \sum_{\mu=0}^{3} \big( \sigma_x^\mu \sigma_x^\mu \partial_{x^\mu} \partial_{x^\mu} J_I(\tau, \mathbf x, \mathbf y) + 2 \epsilon \eta^{\mu \mu} \sigma_x^\mu \sigma_y^\mu \partial_{x^\mu} \partial_{y^\mu} J_I(\tau, \mathbf x, \mathbf y) + \sigma_y^\mu \sigma_y^\mu \partial_{y^\mu} \partial_{y^\mu} J_I(\tau, \mathbf x, \mathbf y) \big).
\end{aligned}
\end{equation}

To find the optimal controls $\mathbf{v}$ and $\mathbf{u}$ from Eq.~\eqref{eq:HJB_R}, we need to take the derivative of the function inside the $\min$ operator with respect to $v^\mu$ and $u^\mu$ and set them to zero. 
This is the standard approach for finding the minimum of a multivariable function, where the critical points are determined by setting the partial derivatives equal to zero.

The optimal control policies $\mathbf v$ and $\mathbf u$, which minimize the expected cost, can be determined from the following conditions:
\begin{equation}
\label{eq:min_u_v_deriv_re}
\begin{aligned}
\partial_{v^\mu} \text{Re} \big(\mathcal{L}(w^\mu, \mathbf z) \big) + \partial_{x^\mu} J_R (\tau, \mathbf x, \mathbf y) = 0,\\
\partial_{u^\mu} \text{Re} \big(\mathcal{L}(w^\mu, \mathbf z) \big) + \partial_{y^\mu} J_R (\tau, \mathbf x, \mathbf y) = 0.
\end{aligned}
\end{equation}

For the imaginary part of the action, the minimization conditions are obtained from:
\begin{equation}
\label{eq:min_u_v_deriv_im}
\begin{aligned}
\partial_{v^\mu} \text{Im} \big(\mathcal{L}(w^\mu, \mathbf z) \big) + \partial_{x^\mu} J_I (\tau, \mathbf x, \mathbf y) = 0,\\
\partial_{u^\mu} \text{Im} \big(\mathcal{L}(w^\mu, \mathbf z) \big) + \partial_{y^\mu} J_I (\tau, \mathbf x, \mathbf y) = 0. \\
\end{aligned}
\end{equation}

These conditions ensure that the stochastic HJB equations for both the real and imaginary actions are satisfied, leading to the minimum expected value of the action across all possible trajectories of the system.
Later, we will prove that the equations derived for the optimal control policies, specifically equations~\eqref{eq:min_u_v_deriv_re} and~\eqref{eq:min_u_v_deriv_im}, are equivalent. 

By multiplying equation~\eqref{eq:HJB_I} by the imaginary unit $i$ and adding it to
equation~\eqref{eq:HJB_R}, we obtain the combined formal form of the stochastic HJB equation for the complex action.
\begin{equation}
\begin{aligned}
\label{eq:group_temrs_in_HJB}
& -\partial_\tau J(\tau, \mathbf x, \mathbf y) = \\
& =\min_{\mathbf v, \mathbf u} \Big( \mathcal{L}(\tau, \mathbf x, \mathbf y, \mathbf v, \mathbf u) + \\
& \quad + \sum_{\mu=0}^{3} \big( v^\mu \partial_{x^\mu} J_R(\tau, \mathbf x, \mathbf y) + u^\mu \partial_{y^\mu} J_R(\tau, \mathbf x, \mathbf y) + i v^\mu \partial_{x^\mu} J_I(\tau, \mathbf x, \mathbf y) + i u^\mu \partial_{y^\mu} J_I(\tau, \mathbf x, \mathbf y) \big) \Big) + \\
& \quad + \frac{1}{2} \sum_{\mu=0}^{3} \big( \sigma_x^\mu \sigma_x^\mu \partial_{x^\mu} \partial_{x^\mu} J_R(\tau, \mathbf x, \mathbf y) + 2 \epsilon \eta^{\mu \mu} \sigma_x^\mu \sigma_y^\mu \partial_{x^\mu} \partial_{y^\mu} J_R(\tau, \mathbf x, \mathbf y) + \sigma_y^\mu \sigma_y^\mu \partial_{y^\mu} \partial_{y^\mu} J_R(\tau, \mathbf x, \mathbf y) \big) + \\
& \quad + \frac{1}{2} \sum_{\mu=0}^{3} \big( i \sigma_x^\mu \sigma_x^\mu \partial_{x^\mu} \partial_{x^\mu} J_I(\tau, \mathbf x, \mathbf y) + 2 i \epsilon \eta^{\mu \mu} \sigma_x^\mu \sigma_y^\mu \partial_{x^\mu} \partial_{y^\mu} J_I(\tau, \mathbf x, \mathbf y) + i \sigma_y^\mu \sigma_y^\mu \partial_{y^\mu} \partial_{y^\mu} J_I(\tau, \mathbf x, \mathbf y) \big) \\
&= \min_{\mathbf v, \mathbf u} \Big( \mathcal{L}(\tau, \mathbf x, \mathbf y, \mathbf v, \mathbf u) + \\
& \quad + \sum_{\mu=0}^{3} \big( v^\mu (\partial_{x^\mu} J_R(\tau, \mathbf x, \mathbf y) + i \partial_{x^\mu} J_I(\tau, \mathbf x, \mathbf y) ) + i u^\mu ( \partial_{y^\mu} J_I(\tau, \mathbf x, \mathbf y) - i \partial_{y^\mu} J_R(\tau, \mathbf x, \mathbf y) ) \big) \Big)+ \\
& \quad + \frac{1}{2} \sum_{\mu=0}^{3} \big(\sigma_x^\mu \sigma_x^\mu ( \partial_{x^\mu} \partial_{x^\mu} J_R(\tau, \mathbf x, \mathbf y) + i \partial_{x^\mu} \partial_{x^\mu} J_I(\tau, \mathbf x, \mathbf y) ) \big) + \\
& \quad + \frac{1}{2} \sum_{\mu=0}^{3} \big(- \sigma_y^\mu \sigma_y^\mu (- \partial_{y^\mu} \partial_{y^\mu} J_R(\tau, \mathbf x, \mathbf y) - i \partial_{x^\mu} \partial_{y^\mu} J_I(\tau, \mathbf x, \mathbf y) ) \big) + \\
& \quad + \frac{1}{2} \sum_{\mu=0}^{3} \big(2 i \epsilon \eta^{\mu \mu} \sigma_x^\mu \sigma_y^\mu (\partial_{x^\mu} \partial_{y^\mu} J_I(\tau, \mathbf x, \mathbf y) - i \partial_{x^\mu} \partial_{y^\mu} J_R(\tau, \mathbf x, \mathbf y) ) \big).
\end{aligned}
\end{equation}

In the above equation~\eqref{eq:group_temrs_in_HJB} we introduce the following notation for the complex minimum.
When the operator $\min$ is applied to a complex function, we refer to this operator as the complex minimum, which we define using the following equation:
\begin{equation}
\label{eq:complex_min}
A(x,y) + i B(x,y) = \min_{v,u} \big(C(x,y,v,u) + i D(x,y,v,u) \big) \iff
\begin{aligned}
 & \left\{
  \begin{aligned}
   A(x,y) &= \min_{v,u} \big(C(x,y,v,u) \big) \\
   B(x,y) &= \min_{v,u} \big(D(x,y,v,u) \big)
  \end{aligned}
 \right.,
\end{aligned}
\end{equation}
note that this notation is only applicable if the $u$ and $v$ minimize both functions $C$ and $D$ simultaneously.
Later, we will prove that this condition is satisfied for the HJB equation, which justifies the correctness of the above notation.

Since the complex intermediate action $J(\tau, \mathbf z)$ is analytic, the equation below explicitly defines its complex derivative:
\begin{equation}
\label{eq:complex_derivative}
\partial_{z^\mu} J(\tau, \mathbf z) = \partial_{x^\mu} J_R(\tau, \mathbf x, \mathbf y) + i \partial_{x^\mu} J_I(\tau, \mathbf x, \mathbf y) = \partial_{y^\mu} J_I(\tau, \mathbf x, \mathbf y) - i \partial_{y^\mu} J_R(\tau, \mathbf x, \mathbf y)
\end{equation}

Furthermore, the Cauchy-Riemann equations, essential for confirming the analyticity of the function, are given by:
\begin{equation}
\label{eq:couchy-rieman}
\partial_{x^\mu} J_R(\tau, \mathbf x, \mathbf y)= \partial_{y^\mu} J_I (\tau, \mathbf x, \mathbf y), \quad \partial_{x^\mu} J_I(\tau, \mathbf x, \mathbf y) = - \partial_{y^\mu} J_R(\tau, \mathbf x, \mathbf y).
\end{equation}

The second complex derivative of the intermediate action is expressed as follows:
\begin{equation}
\begin{aligned}
\partial_{z^\mu} \partial_{z^\mu} J(\tau, \mathbf z) & = \\
& = \partial_{x^\mu} \partial_{x^\mu} J_R(\tau, \mathbf x, \mathbf y) + i \partial_{x^\mu} \partial_{x^\mu} J_I(\tau, \mathbf x, \mathbf y) \\
& = - \partial_{y^\mu} \partial_{y^\mu} J_R(\tau, \mathbf x, \mathbf y) - i \partial_{y^\mu} \partial_{y^\mu} J_I(\tau, \mathbf x, \mathbf y) \\
& = \partial_{x^\mu} \partial_{y^\mu} J_I(\tau, \mathbf x, \mathbf y) - i \partial_{x^\mu} \partial_{y^\mu} J_R(\tau, \mathbf x, \mathbf y),
\qquad \mu=0,1,2,3.
\end{aligned}
\end{equation}

Using the above equations, we can simplify the HJB equation:
\begin{equation}
\begin{aligned}
& -\partial_\tau J(\tau, \mathbf z)  = \\
& = \min_{\mathbf w} \Big( \mathcal{L}(\tau, \mathbf z, \mathbf w) + \sum_{\mu=0}^{3} w^\mu \partial_{z^\mu} J(\tau, \mathbf z) \Big) + \frac{1}{2} \sum_{\mu=0}^{3} \big(\sigma_x^\mu \sigma_x^\mu - \sigma_y^\mu \sigma_y^\mu + 2 i \epsilon \eta^{\mu \mu} \sigma_x^\mu \sigma_y^\mu \big) \partial_{z^\mu} \partial_{z^\mu} J(\tau, \mathbf z) \\
& = \min_{\mathbf w} \Big( \mathcal{L}(\tau, \mathbf z, \mathbf w) + \sum_{\mu=0}^{3} w^\mu \partial_{z^\mu} J(\tau, \mathbf z) \Big) + \frac{1}{2} \sum_{\mu=0}^{3} \sigma^\mu \sigma^\mu \partial_{z^\mu} \partial_{z^\mu} J(\tau, \mathbf z).
\end{aligned}
\end{equation}

Finally, we can formally write the complex HJB equation:
\begin{equation}
\label{eq:HJB_Z}
- \partial_\tau J(\tau, \mathbf z)=\min_{\mathbf w} \Big(\mathcal{L}(\tau, \mathbf z, \mathbf w) + w^\mu \partial_{z^\mu} J(\tau, \mathbf z) \Big) + \frac{1}{2} \sum_{\mu=0}^{3} \sigma^\mu \sigma^\mu \partial_{z^\mu} \partial_{z^\mu} J(\tau, \mathbf z),
\end{equation}
where the complex diffusion coefficient satisfies:
\begin{equation}
\label{eq:complex_diffusion}
\sigma^\mu \sigma^\mu = \sigma_x^\mu \sigma_x^\mu - \sigma_y^\mu \sigma_y^\mu + 2 i \epsilon \eta^{\mu \mu} \sigma_x^\mu \sigma_y^\mu, \quad \mu=0,1,2,3.
\end{equation}

 
Note that, formally, the stochastic increments of perfectly correlated or anti-correlated processes can be expressed as: $dW^\mu=dW^\mu_x=\epsilon \eta^{\mu \mu} dW^\mu_y$. 
Using equation~\eqref{eq:Scaled_Complex_Stochastic_Increments}, we can express the complex diffusion term as: $\sigma^\mu dW^\mu = \sigma^\mu_x dW^\mu + i \epsilon \eta^{\mu \mu} \sigma^\mu_y dW^\mu = ( \sigma^\mu_x  + i \epsilon \eta^{\mu \mu} \sigma^\mu_y) dW^\mu$. From this expression, it follows that the complex diffusion coefficient is: $\sigma^\mu = \sigma^\mu_x + i \epsilon \eta^{\mu \mu} \sigma^\mu_y$, which is in agreement with equation~\eqref{eq:complex_diffusion}.

It is crucial to emphasize again that this formal form of the HJB equation is conceptually meaningful only when considering the distinct equations for its real ~\eqref{eq:HJB_R} and imaginary ~\eqref{eq:HJB_I} parts.

It is clear from its definition that the boundary condition for $J(\tau, \mathbf z)$ is:
\begin{equation}
\label{eq:boundary_condition}
J(\tau_\mathrm{f}, \mathbf z_{\tau_\mathrm{f}}) = 0.
\end{equation}

Sinces $\mathcal{L}(\mathbf z, \mathbf w)$ is an analytic function, the derivative operator and the operator for taking the real part commute. 
Consequently, equation~\eqref{eq:min_u_v_deriv_re} can be expressed as:
\begin{equation}
\label{eq:min_u_v_re}
\begin{aligned}
\text{Re}(\partial_{v^\mu} \mathcal{L}(\mathbf z, \mathbf w)) + \partial_{x^\mu} J_R (\tau, \mathbf x, \mathbf y) = 0, \\
\text{Re}(\partial_{u^\mu} \mathcal{L}(\mathbf z, \mathbf w)) + \partial_{y^\mu} J_R (\tau, \mathbf x, \mathbf y) = 0. \\
\end{aligned}
\end{equation}

Similarly, from equation~\eqref{eq:min_u_v_deriv_im}, we can derive the equations for the optimal control that result from minimizing the imaginary part of the intermediate action:
\begin{equation}
\label{eq:min_u_v_imag}
\begin{aligned}
\text{Im}(\partial_{v^\mu} \mathcal{L}(\mathbf z, \mathbf w)) + \partial_{x^\mu} J_I (\tau, \mathbf x, \mathbf y) = 0, \\
\text{Im}(\partial_{u^\mu} \mathcal{L}(\mathbf z, \mathbf w)) + \partial_{y^\mu} J_I (\tau, \mathbf x, \mathbf y) = 0. \\
\end{aligned}
\end{equation}

We will prove that Equations~\eqref{eq:min_u_v_re} and~\eqref{eq:min_u_v_imag} are equivalent. To do this, we find the derivatives:
\begin{equation}
\begin{aligned}
\partial_{v^\mu} \mathcal{L}(\mathbf z, \mathbf w) = \partial_{w^\mu} \mathcal{L}(\mathbf z, \mathbf w) \frac{\partial w^\mu}{\partial v^\mu}= \partial_{w^\mu} \mathcal{L}(\mathbf z, \mathbf w), \\
\partial_{u^\mu} \mathcal{L}(\mathbf z, \mathbf w) = \partial_{w^\mu} \mathcal{L}(\mathbf z, \mathbf w) \frac{\partial w^\mu}{\partial u^\mu}= i \partial_{w^\mu} \mathcal{L}(\mathbf z, \mathbf w). \\
\end{aligned}
\end{equation}

If we substitute the above derivatives into Equation~\eqref{eq:min_u_v_re}, multiply the second equation by the imaginary unit, and substract it from the first equation, we obtain:
\begin{equation}
\begin{aligned}
\text{Re}(\partial_{w^\mu} \mathcal{L}(\mathbf z, \mathbf w)) - i \text{Re}(i \partial_{w^\mu} \mathcal{L}(\mathbf z, \mathbf w)) + \partial_{x^\mu} J_R (\tau, \mathbf x, \mathbf y) - i \partial_{y^\mu} J_R (\tau, \mathbf x, \mathbf y) = 0.
\end{aligned}
\end{equation}

Similarly, if we substitute the above derivatives into Equation~\eqref{eq:min_u_v_imag}, multiply the first equation by the imaginary unit, and add it to the second equation, we obtain:
\begin{equation}
\begin{aligned}
i \text{Im}(\partial_{w^\mu} \mathcal{L}(\mathbf z, \mathbf w)) + \text{Im}(i \partial_{w^\mu} \mathcal{L}(\mathbf z, \mathbf w)) + i \partial_{x^\mu} J_I (\tau, \mathbf x, \mathbf y) + \partial_{y^\mu} J_I (\tau, \mathbf x, \mathbf y) = 0.
\end{aligned}
\end{equation}

From the Cauchy-Riemann equation~\eqref{eq:couchy-rieman}, the definition of the complex derivative~\eqref{eq:complex_derivative}, and the identities $Z=\text{Re}(Z)-i \text{Re} (i Z)$ and $Z=\text{Im} (i Z) + i \text{Im}(Z)$, where $Z$ is any complex number, we prove that both equations can be written as:
\begin{equation}
\label{eq:complex_optimal_control}
\begin{aligned}
\partial_{w^\mu} \mathcal{L}(\mathbf z, \mathbf w) + \partial_{z^\mu} J (\tau, \mathbf z) = 0.
\end{aligned}
\end{equation}

The equation for the complex optimal control policy $\mathbf w$ maintains the same form as that of the real-valued HJB equation. However, it requires taking a complex derivative of the intermediate action, as can be observed.

Once we identify the complex optimal control $w^*_\mu$ that satisfies equation~\eqref{eq:complex_optimal_control} and substitute it into the complex stochastic HJB equation~\eqref{eq:HJB_Z}, we can eliminate the minimization function. 
To determine the validity of this operation, it is important to remember that the complex stochastic HJB equation is conceptually meaningful only when considering the distinct equations for its real (equation~\eqref{eq:HJB_R}) and imaginary (equation~\eqref{eq:HJB_I}) components. 
By substituting the real $v^*_\mu=\text{Re}(w^*_\mu)$ and the imaginary $u^*_\mu=\text{Im}(w^*_\mu)$ parts of the optimal control into equations~\eqref{eq:HJB_R} and~\eqref{eq:HJB_I} respectively, this allows us to remove the minimization from these equations. 
After integrating both equations into a single complex equation, one can verify that the same result is achieved if we formally replace the complex optimal control in equation~\eqref{eq:HJB_particle_in_EM}.

In the following section, we will illustrate this concept through a specific example by applying a concrete Lagrangian to a relativistic particle in an electromagnetic field.

\section{Analytic continuation of the Covariant Relativistic Lagrangian}
\label{sec:covariant_relativistic_lagrangian}
As an illustration and application of Complex Stochastic Optimal Control theory to a simple relativistic quantum system, we consider the relativistic Lagrangian for a particle in an electromagnetic field, given by:
\begin{equation}
\label{eq:sqrt_lagrangian_real}
\Lambda= \tilde{\sigma} m c \sqrt{\tilde{\sigma} \sum_{\mu=0}^{3} v^\mu v_\mu} + q \sum_{\mu=0}^{3} A_\mu v^\mu,
\end{equation}
where $q$ represents the charge of the particle and $A_\mu$ denotes the 4-vector potential. The symbol $\tilde{\sigma}$ indicates the sign convention for the metric tensor: it takes the value of $+1$ for the metric with diagonal elements $(1,-1,-1,-1)$ and $-1$ for the metric $(-1,1,1,1)$, as elaborated in~\cite{Brizard2009}.

The components of the four-velocity of the particle are related to the speed of light by the equation:

\begin{equation}
\label{eq:weak_equation}
\sum_{\mu=0}^{3} v^\mu v_\mu = \tilde{\sigma} c^2.
\end{equation}

This relation, referred to as the ``weak equation'' by Dirac, allows us to treat $v^\mu$ as unconstrained quantities until all differentiation operations have been carried out, at which point we impose the condition of equation~\eqref{eq:weak_equation} (see~\cite{Goldstein2002} Chapter 7.10).
This will be the approach we employ as we seek to minimize the expected value of the stochastic action.

The Lagrangian in equation~\eqref{eq:sqrt_lagrangian_real} is a real-valued function of real arguments -- the coordinates and velocity of the particle. In complex stochastic optimal control, we assume that this Lagrangian is the analytic continuation of the real-valued Lagrangian referenced in equation~\eqref{eq:sqrt_lagrangian_real}.
\begin{equation}
\label{eq:sqrt_lagrangian}
\mathcal{L}(\mathbf z, \mathbf w)= \tilde{\sigma} m c \sqrt{\tilde{\sigma} \sum_{\mu=0}^{3} w^\mu w_\mu} + q \sum_{\mu=0}^{3} A_\mu(\tau, \mathbf z) w^\mu.
\end{equation}

The ``weak equation'' should be also analytically continued:
\begin{equation}
\label{eq:w_weak_equation}
\sum_{\mu=0}^{3} w^\mu w_\mu = \tilde{\sigma} c^2.
\end{equation}

The derivative of the complex Lagrangian can be calculated using the ``weak equation''~\eqref{eq:w_weak_equation}:
\begin{equation}
\label{eq:lagrangian_derivative}
\begin{aligned}
\partial_{w^\mu} \mathcal{L}(\mathbf z, \mathbf w) &=\partial_{w^\mu} \Bigg(\tilde{\sigma} m c \sqrt{\tilde{\sigma} \sum_{\mu=0}^{3} w^\mu w_\mu} + q \sum_{\mu=0}^{3} A_\mu(\tau, \mathbf z) w^\mu \Bigg) =\\
& =\frac{2 \tilde{\sigma}^2 mc w_\mu }{2 \sqrt{\tilde{\sigma} \sum_{\mu=0}^{3} w^\mu w_\mu} } + q A_\mu(\tau, \mathbf z) = m w_\mu + q A_\mu(\tau, \mathbf z).
\end{aligned}
\end{equation}

Finally, by substituting Equation~\eqref{eq:lagrangian_derivative} into Equation~\eqref{eq:complex_optimal_control}, we obtain the expression for the complex optimal control:
\begin{equation}
\begin{aligned}
\label{eq:complex_opitmal_control_particle_in_EM}
w^*_\mu = -\frac{1}{m} (\partial_{z^\mu} J (\tau, \mathbf z) + q A_\mu(\tau, \mathbf z)).
\end{aligned}
\end{equation}

If we substitute the analytically continued Lagrangian from Equation~\eqref{eq:sqrt_lagrangian} into the complex HJB equation referenced in Equation~\eqref{eq:HJB_Z}, we obtain:
\begin{equation}
\label{eq:HJB_particle_in_EM}
- \partial_\tau J(\tau, \mathbf z)=\min_{\mathbf w} \Bigg(\tilde{\sigma} m c \sqrt{\tilde{\sigma} \sum_{\mu=0}^{3} w^\mu w_\mu} + \sum_{\mu=0}^{3} w^\mu \big(\partial_{z^\mu} J(\tau, \mathbf x) + q A_\mu(\tau, \mathbf z) \big) \Bigg)+ \frac{1}{2} \sum_{\mu=0}^{3} \sigma^\mu \sigma^\mu \partial_{z^\mu} \partial_{z^\mu} J(\tau, \mathbf z).
\end{equation}

As noted at the end of Section~\ref{sec:complex_HJB}, we can formally remove the minimization function from the above equation by substituting in the complex optimal control policy from equation~\eqref{eq:complex_opitmal_control_particle_in_EM}. This demonstrates the validity of the formal approach to the complex stochastic HJB equations used to derive the Dirac equation in~\cite{Yordanov2024}.

\section{Stochastic equation of motion for quantum particle}
\label{sec:stochastic_equation_of_motion}
In Section~\ref{sec:complex_HJB}, we derived equation~\eqref{eq:complex_diffusion} for the complex diffusion coefficient. 
We would like to emphasize that the imaginary term in this equation results from the assumption of perfect correlation or ant-correlation (correlation coefficient of $\pm 1$) between the stochastic processes $W^\mu_x$ and $W^\mu_y$. 
The concept of correlating the real and imaginary stochastic processes is similar to that introduced in~\cite{Kuipers2023}.

Let us postulate that the real and imaginary diffusion coefficients are equal, and are given by:
\begin{equation}
\label{eq:sigma_x_sigma_y}
\sigma_x^\mu \sigma_x^\mu=\sigma_y^\mu \sigma_y^\mu=\frac{\hbar}{m}, \qquad \mu=0,1,2,3,
\end{equation}
note that a similar diffusion coefficient was used by Nelson~\cite{Nelson1966}, though he considered forward and backward stochastic processes in time.

Substituting equation~\eqref{eq:sigma_x_sigma_y} into equation~\eqref{eq:complex_diffusion}, we obtain that the complex diffusion coefficient is purely imaginary:
\begin{equation}
\label{eq:sigma_sigma}
\sigma^\mu \sigma^\mu=2 i \epsilon \eta^{\mu \mu} \frac{\hbar}{m}, \qquad \mu=0,1,2,3,
\end{equation}
where $\epsilon=+1$ if the real and imaginary stochastic processes are perfectly positively correlated, and $\epsilon=-1$ if these processes are perfectly anti-correlated.
Note that the presence of the metric tensor components $\eta^{\mu \mu}$ in equation~\eqref{eq:sigma_sigma} preserve the covariant form of the second order term in the HJB Equation~\eqref{eq:HJB_particle_in_EM}. 
By substituting equation~\eqref{eq:sigma_sigma} into the last term of equation~\eqref{eq:HJB_particle_in_EM} we obtain: $\frac{1}{2} \sum_{\mu=0}^{3} \sigma^\mu \sigma^\mu \partial_{z^\mu} \partial_{z^\mu} J(\tau, \mathbf z) = i \epsilon \frac{\hbar}{m} \sum_{\mu=0}^{3} \eta^{\mu \mu} \partial_{z^\mu} \partial_{z^\mu} J(\tau, \mathbf z) = i \epsilon \frac{\hbar}{m} \sum_{\mu=0}^{3} \partial^{z^\mu} \partial_{z^\mu} J(\tau, \mathbf z) $. This demonstrates that the term is proportional to the d’Alembert operator acting on $J$, which is a Lorentz-invariant operator, confirming the Lorentz covariance of the HJB equation.

This leads to the stochastic equations of motion for the particle:
\begin{equation}
d x^\mu=v^\mu ds + \sqrt{\frac{\hbar}{m}} dW^\mu_x, \qquad \mu=0,1,2,3,
\end{equation}

\begin{equation}
d y^\mu=u^\mu ds + \sqrt{\frac{\hbar}{m}} dW^\mu_y, \qquad \mu=0,1,2,3.
\end{equation}

Considering the correlation properties of the stochastic processes $W^\mu_x$ and $W^\mu_y$ discussed in the previous section, we can express these equations as:
\begin{equation}
d x^\mu=v^\mu ds + \sqrt{\frac{\hbar}{m}} dW^\mu, \qquad \mu=0,1,2,3,
\end{equation}

\begin{equation}
d y^\mu=u^\mu ds + \epsilon \eta^{\mu \mu} \sqrt{\frac{\hbar}{m}} dW^\mu, \qquad \mu=0,1,2,3.
\end{equation}

In our previous work~\cite{Yordanov2024} (see also Appendix~\ref{sec:linearizing_HJB} for more details), we demonstrated that to linearize the HJB equation, the diffusion coefficient must satisfy the above Equation~\eqref{eq:sigma_sigma} for the diffusion coefficient.

This result demonstrates that the HJB equation can be linearized to obtain the Dirac equation, not by assuming the specific form of the diffusion coefficient in equation~\eqref{eq:sigma_sigma}, but by assuming perfect correlation of the real and imaginary stochastic processes that describe the particle's complex stochastic equation of motion.

Another observation is that if perfect correlation is assumed for a given four-coordinate component of particles, then perfect anti-correlation is required for the corresponding component of anti-particles.

\section{Complex Motion of Particles}
\label{sec:complex_stochastic_motion}
There are some recent weak measurement experiments~\cite{Aharonov1988,Kocsis2011,Mahler2016,Murch2013,Rossi2019,Hariri2019,Wang2022} that motivate an objective description of a quantum system in the time interval between two complete measurements in terms of two state vectors, together with a new type of physical quantity, the complex “weak value” of a quantum mechanical observable~\cite{Aharonov2005,Jozsa2007,Mori2015}.

The real part of the weak value may correspond to an average over these stochastic trajectories, representing the expected outcome of the weak measurement. On the other hand, the imaginary part of the weak value could be associated with fluctuations or uncertainties in the stochastic paths~\cite{Aharonov2005}.

In Stochastic Mechanics, the weak value can be interpreted as an effective statistical quantity arising from averaging over the many possible complex stochastic paths between the pre-selected and post-selected state vectors.

The outcome of these weak measurement experiments is a complex value. This might suggest the existence of complex coordinates for particles, or Complex Stochastic Mechanics can be recognized as a powerful mathematical framework for describing these experiments at the microscopic level.

\section{Conclusion}
In this paper, we have demonstrated that real variables, such as the four-coordinates, control policy, and cost-to-go function in the HJB equation, can be formally substituted with complex-valued counterparts. 
However, this substitution is valid only if the complex-valued Lagrangian is defined as an analytical continuation of the system’s real-valued Lagrangian. 
Additionally, the complex stochastic motion of the particles must be characterized by two correlated stochastic processes: one governing the real coordinates and the other governing the imaginary coordinates of the particle.

We provided a rigorous mathematical procedure for minimizing the expected value of the complex stochastic action of the particle. We showed that the formal structure of the complex HJB equation, as well as the equation for the velocity field (i.e., the optimal control policy), remains consistent with that of the real-valued case.

By applying the Cauchy-Riemann theorem to the intermediate action (the cost-to-go function) $J$, and assuming perfect correlation between the real and imaginary stochastic processes, we derived the complex diffusion coefficient of the complex stochastic process. 
To preserve the covariant form of the HJB equation, we demonstrated that if the temporal parts of the two stochastic processes are perfectly correlated, then their spatial parts must be perfectly anti-correlated.

In contrast to our previous work, where the derivation of the complex diffusion coefficient was based on the requirements for linearizing the HJB equation, in this paper, we derived the same diffusion coefficient from more fundamental principles.
This newly derived diffusion coefficient, when substituted into the complex stochastic HJB equation, allows the HJB equation to be linearized, thereby leading to the derivation of the equations of relativistic quantum mechanics.

\section*{Acknowledgments}
I wish to express my gratitude to Folkert Kuipers for the valuable discussions. He also pointed out the need to consider the perfect correlation of the real and imaginary stochastic terms. 
I would also like to thank Markku Karhunen for his useful remarks and critical comments on a previous version of this paper, as well as for verifying the mathematical correctness of all equations in that version, which significantly contributed to the refinement of this manuscript.

\newpage
\appendix
\section{Appendix: Linearization of stochastic Hamilton-Jacobi-Belman equation for a particle in electromagnetic field}
\label{sec:linearizing_HJB}
In this appendix, we will linearize the stochastic HJB equation for a particle in an electromagnetic field, as derived in Section~\ref{sec:covariant_relativistic_lagrangian}.
\begin{equation}
\label{eq:HJB_particle_in_EM2}
- \partial_\tau J(\tau, \mathbf z)=\min_{\mathbf w} \Bigg(\tilde{\sigma} m c \sqrt{\tilde{\sigma} \sum_{\mu=0}^{3} w^\mu w_\mu} + \sum_{\mu=0}^{3} w^\mu \big( q A_\mu(\tau, \mathbf z) + \partial_{z^\mu} J(\tau, \mathbf z) \big) \Bigg) + \frac{1}{2} \sum_{\mu=0}^{3} \sigma^\mu \sigma^\mu \partial_{z^\mu} \partial_{z^\mu} J(\tau, \mathbf z).
\end{equation}

In the same Section~\ref{sec:covariant_relativistic_lagrangian}, we derived the equation for the optimal control:
\begin{equation}
\begin{aligned}
\label{eq:complex_opitmal_control_particle_in_EM2}
w^*_\mu = -\frac{1}{m} (\partial_{z^\mu} J (\tau, \mathbf z) + q A_\mu(\tau, \mathbf z)).
\end{aligned}
\end{equation}

In Section~\ref{sec:stochastic_equation_of_motion}, we derived the equation for the complex diffusion coefficient:
\begin{equation}
\label{eq:sigma_sigma2}
\sigma^\mu \sigma^\mu=2 i \epsilon \eta^{\mu \mu} \frac{\hbar}{m}, \qquad \mu=0,1,2,3,
\end{equation}

As pointed out in our previous work~\cite{Yordanov2024}, to linearize the stochastic HJB equation, we use a similar approach to that which Dirac~\cite{Dirac1928} employed to linearize the relativistic energy.
\begin{equation}
\label{eq:lagrangian_linearization}
\tilde{\sigma} m c \sqrt{\tilde{\sigma} \sum_{\mu=0}^{3} w^\mu w_\mu} = m c \gamma^\mu w_\mu
\end{equation}

If we substitute Equations~\eqref{eq:complex_opitmal_control_particle_in_EM2},\eqref{eq:sigma_sigma2}, and\eqref{eq:lagrangian_linearization} into the stochastic HJB Equation~\eqref{eq:HJB_particle_in_EM2}, and remove the minimization function since the used four-velocity is the optimal one, we obtain:
\begin{equation}
\begin{aligned}
& - \partial_\tau \bm{J}^{(r)}(\tau, \mathbf z)= \\
& = - \epsilon_r m c \sum_{\mu=0}^{3} \gamma^\mu \big( \partial_{z^\mu} \bm{J}^{(r)}(\tau, \mathbf z) + q \bm{A}_\mu(\tau, \mathbf z) \big) + 
 \frac{1}{2} \sum_{\mu=0}^{3} 2 i \epsilon_r \eta^{\mu \mu} \frac{\hbar}{m} \partial_{z^\mu} \partial_{z^\mu} \bm{J}^{(r)}(\tau, \mathbf z) + \\
& \quad + \sum_{\mu=0}^{3} \big( \partial^{z^\mu} \bm{J}^{(r)}(\tau, \mathbf z) + q A^\mu(\tau, \mathbf z) \big) \big( \partial_{z^\mu} \bm{J}^{(r)}(\tau, \mathbf z) + q A_\mu(\tau, \mathbf z) \big), \quad r=1,2,3,4.
\end{aligned}
\end{equation}
Here, $\bm{J}(\tau, \mathbf z)=J(\tau, \mathbf z) \mathbb{I}_4$, where $\mathbb{I}_4$ is the identity matrix. The $r$-th component of the vector colum $\bm{J}(\tau, \mathbf z)$ is denoted as $\bm{J}^{(r)}(\tau, \mathbf z)$. $\epsilon_r = +1, \text{if } r=1,2$ and $\epsilon_r = -1, \text{if } r=3,4$ 

Expanding the last term of the above equation, we obtain:
\begin{equation}
\begin{aligned}
\label{eq:HJB_J_expanded}
& - \partial_\tau \bm{J}^{(r)}(\tau, \mathbf z)= \\
&=- \epsilon_r mc \sum_{\mu=0}^{3} \gamma^\mu \frac{1}{m} \left(\partial_\mu \bm{J}^{(r)}(\tau, \mathbf z) + q A_\mu \right) + \frac{i \epsilon_r \hbar}{m} \sum_{\mu=0}^{3} \partial^\mu \partial_\mu \bm{J^{(r)}}(\tau, \mathbf z) -\\
& \quad - \frac{1}{m} \sum_{\mu=0}^{3} \partial^\mu \bm{J}^{(r)}(\tau, \mathbf z) \partial_\mu \bm{J}^{(r)}(\tau, \mathbf z) - 2 \frac{1}{m} \sum_{\mu=0}^{3} \partial^\mu \bm{J}^{(r)}(\tau, \mathbf z) q A_\mu - \frac{1}{m} q^2 \sum_{\mu=0}^{3} A^\mu A_\mu, \, r=1,2,3,4.
\end{aligned}
\end{equation}

Let us introduce a new function, $\tilde{J}(\tau, \mathbf z)$:
\begin{equation}
\label{eq:S_definition}
\bm{J}^{(r)}(\tau, \mathbf z) = -i \epsilon_r \hbar \tilde{J}(\tau, \mathbf z) \mathbb{I}^{(r)}_4, \quad r=1,2,3,4.
\end{equation}

Substituting Equation~\eqref{eq:S_definition} into Equation~\eqref{eq:HJB_J_expanded}, and then multiplying both sides by $m$, we obtain:
\begin{equation}
\begin{aligned}
& i \hbar m \partial_\tau \tilde{J}(\tau, \mathbf z) = \\
&= m c \sum_{\mu=0}^{3} \gamma^\mu \left(i \hbar \partial_\mu \tilde{J}(\tau, \mathbf z) - e A_\mu \right) + \\
& \quad + \hbar^2\left(\sum_{\mu=0}^{3} \partial^\mu \tilde{J}(\tau, \mathbf z) \partial_\mu \tilde{J}(\tau, \mathbf z) + \sum_{\mu=0}^{3} \partial^\mu \partial_\mu \tilde{J}(\tau, \mathbf z) \right) + 2 i \hbar \sum_{\mu=0}^{3} \partial^\mu \tilde{J}(\tau, \mathbf z) e A_\mu - e^2 \sum_{\mu=0}^{3}A^\mu A_\mu.
\end{aligned}
\end{equation}

To linearize the HJB Equation, we are going to use the (Hopf-Cole) logarithmic transformation:
\begin{equation}
\begin{aligned}
& \sum_{\mu=0}^{3} \partial^\mu \tilde{J}(\tau, \mathbf z) \partial_\mu \tilde{J}(\tau, \mathbf z) + \sum_{\mu=0}^{3} \partial^\mu \partial_\mu \tilde{J}(\tau, \mathbf z) = \sum_{\mu=0}^{3} \frac{\partial^\mu \phi(\tau, \mathbf z)}{\phi(\tau, \mathbf z)} \frac{\partial_\mu \phi(\tau, \mathbf z)}{\phi(\tau, \mathbf z)} + \sum_{\mu=0}^{3} \partial^\mu \left(\frac{\partial_\mu \phi(\tau, \mathbf z)}{\phi(\tau, \mathbf z)} \right) = \\
& = \sum_{\mu=0}^{3} \frac{\partial^\mu \phi(\tau, \mathbf z)}{\phi(\tau, \mathbf z)} \frac{\partial_\mu \phi(\tau, \mathbf z)}{\phi(\tau, \mathbf z)} + \sum_{\mu=0}^{3} \frac{\partial^\mu \partial_\mu \phi(\tau, \mathbf z)}{\phi(\tau, \mathbf z)} - \sum_{\mu=0}^{3} \frac{\partial^\mu \phi(\tau, \mathbf z) \partial_\mu \phi(\tau, \mathbf z)}{\phi(\tau, \mathbf z)^2} = \sum_{\mu=0}^{3} \frac{\partial^\mu \partial_\mu \phi(\tau, \mathbf z)}{\phi(\tau, \mathbf z)}.
\end{aligned}
\end{equation}

The above substitution transforms the HJB equation into a linear equation:
\begin{equation}
\begin{aligned}
\label{eq:HJB_divided_by_phi}
& i \hbar m \frac{\partial_\tau \phi(\tau, \mathbf z)}{\phi(\tau, \mathbf z)} = m c \left( i \hbar \gamma^\mu \frac{\partial_\mu \phi(\tau, \mathbf z)}{\phi(\tau, \mathbf z)} - e \gamma^\mu A_\mu \right) + \hbar^2 \frac{\partial^\mu \partial_\mu \phi(\tau, \mathbf z)}{\phi(\tau, \mathbf z)} + 2 e A^\mu i \hbar \frac{\partial_\mu \phi(\tau, \mathbf z)}{\phi(\tau, \mathbf z)} - e^2 A^\mu A_\mu.
\end{aligned}
\end{equation}

Multiplying both sides of Equation~\eqref{eq:HJB_divided_by_phi} by $\phi(\tau, \mathbf{z})$, we finally obtain the linearized HJB equation:
\begin{equation}
\begin{aligned}
\label{eq:hjb_dirac_initial}
& i \hbar m \partial_\tau \phi(\tau, \mathbf z) = \\
&= m c \left( i \hbar \gamma^\mu \partial_\mu \phi(\tau, \mathbf z) - e \gamma^\mu A_\mu \phi(\tau, \mathbf z) \right) + \hbar^2 \partial^\mu \partial_\mu \phi(\tau, \mathbf z) + 2 e A^\mu i \hbar \partial_\mu \phi(\tau, \mathbf z) - e^2 A^\mu A_\mu \phi(\tau, \mathbf z).
\end{aligned}
\end{equation}

One may further consult~\cite{Yordanov2024} to see how the Dirac equation can be derived from the above linearized HJB equation.

\bibliographystyle{unsrt}
\bibliography{refs}

\end{document}